\documentclass[a4paper, 12pt]{article}
\usepackage{graphicx}
\usepackage{amssymb,amsmath}
\usepackage[utf8]{inputenc}  

\title{The radial memory effect in multimode optical fibres}
\author{Ulas~Gokay, Robert~J.~Kilpatrick, Simon~A.R.~Horsley,\\
David~B.~Phillips \& Jacopo~Bertolotti}
\date{Physics and Astronomy Department, University of Exeter, Stocker Road, Exeter EX4 4QL, UK}

\begin{document}

\maketitle

\noindent\textbf{Abstract}

\noindent We present a study of a novel memory effect in multimode optical fibres, which manifests itself as an output ring of excess energy at the same radius as an input focussed spot. This effect is robust against fibre perturbations, and we discuss its possible use for spatial multiplexing.

\section{Introduction}

Optical fibres are a cornerstone of modern information technology, enabling long-distance and high-speed information transfer~\cite{OptCommRoadmap2024}. The Physics of optical fibres is extremely well understood~\cite{okamoto}, but there are still plenty of open technical challenges. Single mode fibres are by far the most commonly used in most applications, but in principle multimode fibre have a higher information capacity, i.e., they can transport a large number of orthogonal spatial modes within a small cross-sectional area. In practice making use of the full capacity of a multimode fibre is difficult due to modal dispersion, meaning that different modes will accumulate different phases when propagating through the fibre, and interfere to form a speckle pattern at the output (as shown in Fig.~\ref{fig:Fig1}(a)), which makes it difficult to disentangle the orthogonal modes ~\cite{gigan2022roadmap,cao2023controlling}.

This problem has been studied in detail in the context of transmitting an image through a multimode fibre~\cite{cizmar2012}, and is compounded by the fact that small perturbations of the fibre (e.g., mild bending) changes the phase velocity of the fibre modes in a mode dependent manner, and can slightly modify the fibre modes, resulting in coupling between modes that were orthogonal in the unperturbed fibre~\cite{Plschner2015}. One could, in principle, solve the problem by fully characterizing the transfer function of the fibre, and thus find the true orthogonal modes, but this is a time-intensive process that typically requires full optical access to both ends of the fibre, and needs to be repeated every time the fibre is perturbed~\cite{Hu2024}. Many different approaches are currently being studied to overcome these problems, from engineering the output facet of a fibre with bespoke reflectors~\cite{gordon2019characterizing,gu2015design} or guide stars~\cite{li2021memory}, to the development of perturbation-resistant fibres~\cite{boonzajer2018robustness}, to the use of machine learning techniques~\cite{resisi2021image, zheng2024self}, to pre-calibrating for a range of possible perturbations~\cite{wen2023single}.

A different approach to the problem is to take advantage of speckle correlations and exploit a \textit{memory effect}~\cite{Feng1988}. Speckle patterns are only random in the sense that their details often depend on poorly controlled variables like the exact position of each scatterer or the specific bending of a fibre, but a given speckle pattern is completely deterministic -- if all variables about a fibre state are known, the details of the speckle pattern can be predicted~\cite{Plschner2015}.
As discussed in more detail in the Supplementary Information, this results in subtle correlations within the random-looking speckle pattern emanating from a multimode fibre~\cite{akkermansbook}. A particularly important class of speckle correlations are the so-called ``memory effects", where changing the input by a small amount results in a predictable change in the output, independently from the fine details of how the system is perturbed. The most famous example in optical fibres is the axial angle conservation in step-index multimode fibres, meaning that illuminating the input of a fibre with a beam at an angle $\theta$ with respect to the optical axis will result in an output largely concentrated in a cone of the same angle, which can be used for angular multiplexing~\cite{AngularMultiplexing1981}. Another, more recently discovered memory effect in fibres is the rotational memory effect, which emerges from the cylindrical symmetry of a fibre, and connects the rotation of the input wavefront to the rotation of the output wavefront~\cite{amitonova2015rotational, gutierrez2024characterization}. These memory effects have recently started to find applications for imaging through multimode fibres~\cite{li2021memory, gomes2025funnelling}.

In this paper we describe a less well-known kind of memory effect, which we dub the \textit{radial memory effect}, which connects the radial position of the input with a ring of extra intensity at the same radius at the output. We characterise the effect and show that it has the potential to be used for spatial multiplexing.

\section{The radial memory effect}
Due to the interference between modes traveling with different phase velocities, the output from a coherent input in a multimode fibre typically looks like a speckle pattern. Changing the incident wavefront excites a different linear combination of fibre modes, thus resulting in a different speckle pattern emerging from the output of the fibre. For an ideal step-index fibre it is possible to derive the mode shapes and their phase velocities~\cite{okamoto} and so one could in principle compute the resulting speckle, but any perturbation of the real system from this ideal model (e.g., bending or tiny changes in length) will change the output speckle in a hard to predict way~\cite{Hu2024,Plschner2015}. Nevertheless, the fibre will still have a complete set of orthogonal modes even if we don't know exactly how to compute them, and an incident wavefront will excite each of them depending on how well it matches them. In particular, illuminating the input fibre facet with a focused spot at a radius $r_0$ from the centre of the fibre core will preferentially couple to the subset of modes with a significant fraction of their electromagnetic field at the same radius.
Assuming mode coupling within the fibre is small, we therefore expect that a focussed input beam will result in an output speckle pattern with an excess of intensity at the same radius $r_0$. We term this phenomenon the \textit{radial memory effect}. We note that these rings of excess intensity have been observed before (e.g.,~\cite{li2021memory,rimoli2024demixing}), but are yet to be explained in detail.

We begin by analysing the radial memory effect theoretically (a full derivation can be found in the Supplementary Information). For an ideal straight segment of step-index multimode fibre, solving the wave equation under the low NA weakly guiding approximation leads to a set of circularly polarised eigenmodes, $\psi_{\ell,p}$, that exhibit minimal polarisation coupling on propagation, so can be treated as scalar fields. Within the step-index core of radius $a$, these eigenmodes can be written in cylindrical coordinates as:
\begin{equation}\label{eqn:eigenmode}
\psi_{\ell,p}\left(r,\theta, z \right)= \frac{1}{N_{\ell,p}}\frac{J_{\ell}\left({u}_{\ell ,p}r/a\right)}{J_{\ell}\left(u_{\ell ,p}\right)} {e}^{{\rm{i}}\ell\theta}\, {e}^{{\rm{i}} \beta_{\ell,p} z},
\end{equation}
where $\ell$ and $p$ enumerate the modes, $J_{\ell}$ is a Bessel function of the first kind of order $\ell$, $u_{\ell ,p}$ is the normalized transverse wave number, $\beta_{\ell,p}$ is the axial ($z$) component of the wavevector (i.e., the global phase change of the mode per metre of propagation along the fibre), and $N_{\ell,p}$ is a normalization factor.

If we assume a sharp focused excitation at position $(r, \theta\, z) = (r_0, \theta_0, 0)$, each mode $\psi_{\ell,p}$ will be excited with a complex coefficient $\sim \psi_{\ell,p}^{\ast}(r_0,\theta_0,0)$. Each mode will then propagate with their own $\beta_{\ell,p}$ and produce an intensity pattern at the end of the fibre ($z=Z$) given by
\begin{equation}
    \label{eq:intensity}
    \begin{split}
    I(r, &\theta, r_0, \theta_0, Z) = \frac{1}{M} \left\lvert \sum_{\ell, p} \psi_{\ell,p}^{\ast}(r_0,\theta_0,0) \, \psi_{\ell,p}(r,\theta, Z) \right\rvert^2 =\\
    &= \frac{1}{M} \sum_{\ell, p} \left\lvert\psi_{\ell,p}^{\ast}(r_0,\theta_0,0) \right\rvert^2 \, \left\lvert \psi_{\ell,p}(r,\theta, Z) \right\rvert^2 +\\
    &+ \frac{1}{M}
    \sum_{\ell \neq\ell^{\prime} \,\text{OR}\, p\neq p^{\prime}}    
    \psi_{\ell,p}^{\ast}(r_0,\theta_0,0) \, \psi_{\ell^{\prime},p^{\prime}}(r_0,\theta_0,0)\, \psi_{\ell,p}(r,\theta, Z)\, \psi_{\ell^{\prime},p^{\prime}}^{\ast}(r,\theta, Z) \, ,
    \end{split}
\end{equation}
where $M$ is the total number of propagating modes in the fibre. Since we are not interested in the detail of the speckle pattern arising from a specific fibre length $Z$, we can average over fibre length (making sure we average over a range of lengths much larger that $\beta_{\ell,p}^{-1}$). We notice that all terms in the second sum on the right hand side of Eqn.~\ref{eq:intensity} vanish unless $\beta_{\ell,p}=\beta_{\ell^{\prime},p^{\prime}}$, which only occurs for fibre modes in which both ${\ell= -\ell^{\prime}}$ and $p=p'$. Thus we find the ensemble averaged output intensity $\left\langle I \right\rangle_Z$ is given by
\begin{equation}
\label{eq:avgint}
    \begin{aligned}
        \left\langle I \right\rangle_Z &= \frac{1}{M}\sum_{\ell=-L}^{L}\sum_{p=0}^{P(\ell)}\frac{J^2_{\ell}(u_{\ell,p}r_0/a)J^2_{\ell}(u_{\ell,p}r/a)}{N_{\ell,p}^4J^{4}_{\ell}(u_{\ell,p})} +\\
        &+ \frac{2}{M}\sum_{\ell=1}^{L}\sum_{p=0}^{P(\ell)}\frac{J^2_{\ell}(u_{\ell,p}r_0/a)J^2_{\ell}(u_{\ell,p}r/a)}{N_{\ell,p}^{4}J^{4}_{\ell}(u_{\ell,p})}\cos(2\ell(\theta-\theta_0)) \, ,
    \end{aligned}
\end{equation}
where $L$ and $P(\ell)$ are, respectively, the positive extreme values of mode indices $\ell$ and $p$.

Figure~\ref{fig:Fig2} shows plots of both Eqn.~\ref{eq:intensity} (at a fixed value of $Z$), and Eqn.~\ref{eq:avgint}, for a several different values of $r_0$. As the Bessel functions $J_{\ell}$ have a well defined maximum and decay rapidly when the argument is smaller than $\ell$, a mode with most of its energy concentrated around $r_0$ will be strongly excited by a focus at $r=r_0$ and vice versa. Therefore, the first term in Eqn.~\ref{eq:avgint} will, especially for larger values of $r_0$, look like a ring of excess energy of radius $r_0$, sitting on top of a relatively flat background (as can be seen in Fig.~\ref{fig:Fig2}(d)). On the contrary, the second term in Eqn.~\ref{fig:Fig2} will be modulated with the angular coordinate, with two maxima at $\theta = \pm \theta_0$, producing a diagonal line of excess energy. It is important to note that the existence of this second term hinges on $\beta_{\ell,p}=\beta_{-\ell,p}$, which easily breaks due to spin-orbit coupling, or when the fibre stops being a perfect cylinder and is bent in any way, thus making the diagonal line feature of $\left\langle I \right\rangle_Z$ unlikely to be observable in experiments.

\section{Experimental measurements}
To explore this phenomenon experimentally, we built an optical system able to image both facets of the multimode fibre and, at the same time illuminate the input end of the fibre with a focused spot at a programmable point $r_0$ on the core.
A schematic of the experimental apparatus is shown in Fig.~\ref{fig:Fig1}(b). Figure~\ref{fig:Fig3} shows typical measurements on a 20\,cm long step-index fibre with a core radius of $a=100\,\mu$m and an NA\,$\simeq0.22$, illuminated with stabilized HeNe laser of wavelength $\lambda=633$\,nm. More data (including data from a second fibre of core radius of $a=25\,\mu$m) can be found in the Supplementary Information.
In all cases, a brighter ring of speckle, at the same radius as the input illumination, is clearly visible at the fibre output.  Averaging over either the illumination position (at constant $r$) or, equivalently, the fibre configuration (by manually shaking the fibre) smooths out the fluctuations due to the speckle statistics, while the ring of excess intensity remains. As expected, we see no evidence of a bright stripe of excess intensity in our experiments. The appearance of this strip relies on reflection symmetry of light totally internally reflecting from opposite sides of the core-cladding boundary. This reflection symmetry is broken in a real fibre. While we did not explore the spectral degree of freedom in this work, we notice that averaging over frequency would also lead to a similar outcome, as implicit in the results shown in~\cite{rimoli2024demixing}.

To confirm the correlation between the input position and the output excess intensity we scan the focused input and plot the output intensity profile projected on the radial coordinate (Figure~\ref{fig:Fig3}, lower panels). For small radial coordinates all the excess intensity is concentrated in a central featureless region, which we do not expect from eq.~\ref{eq:avgint} (it can be seen in Figure~\ref{fig:Fig2} that for small values of $r_0$ we would expect the ring to just have a smaller radius). We attribute this to the greater degree of mode coupling that typically occurs between low-order modes due to the fibre bending, as discussed in refs.~\cite{Plschner2015,wen2023single}.

\section{Spatial multiplexing}
As a ring of excess intensity can be reliably seen at a predictable radius when the fibre is illuminated with a focused beam, and since this effect -- like all memory effects -- is relatively robust against perturbations, we next explore the potential for spatial multiplexing of information. As a proof of principle we looked at a 20~cm long 200$\mu$m core diameter step-index fibre, characterized its average radial response as a function as the input radial position (Figure~\ref{fig:Fig3}), and identified 4 illumination spots corresponding to 4 well-separated output annular regions (see Figure~\ref{fig:Fig4}).

For each of the four possible illuminations we made a calibration measurement by recording the total output intensity in all 4 rings. For a given input, the intensity within the 4 rings is stored in a 4-component column vector. Stacking these column vectors side-by-side results in a $4\times 4$ \textit{response matrix} $\mathcal{M}$, so that $I_{\text{out}}= \mathcal{M} I_{\text{in}}$, where $I_{\text{in}}$ and $I_{\text{out}}$ are both 4-component vectors, representing (respectively) the intensity at the 4 input and output radii. Assuming that nonlinar effects can be neglected (which is a safe assumption, until the fibre becomes very long \cite{Marcuse1991}) knowledge of this response matrix allows us to predict the amount of intensity in the 4 output rings given any linear combination of the 4 possible inputs. More importantly, this matrix is invertible, so we can reconstruct what the linear combination of the input $I_{\text{in}}= \mathcal{M}^{-1}\,I_{\text{out}}$ was just by measuring $I_{\text{out}}$, i.e., the total intensity in each of the 4 rings (e.g., with a camera, like in our proof of principle, or ideally with a diffraction element that sends the light coming from each ring to 4 fast detectors).

To test how well the 4 inputs can be reconstructed we used the system shown in Figure~\ref{fig:Fig1} (and described in detail in section~\ref{section:method}) to send a number of combinations of the 4 possible inputs to the fibre with a Digital Micromirror Device (DMD). In Figure~\ref{fig:Fig4} a selection of typical measurements is shown, comparing the ground truth (which we can measure directly by looking at the reflection from the input facet) and the reconstruction from the transmitted light. We notice that for a practical application one would want to select a threshold and define a signal to be either on or off (i.e., use this system digitally rather than analogically). In our case we see that if we choose a threshold anywhere between 0.2 and 0.3 (in the units used for Figure~\ref{fig:Fig4}) we get a perfect reconstruction.

\section{Discussion and conclusion}
The radial memory effect is a reliable correlation that is robust against perturbations and does not depend on the fine details of the fibre. As it provides partial information of the input wavefront from measuring only the transmitted intensity pattern it opens a novel option for multiplexing data transfer. Interestingly, the radial memory effect exists alongside the more well known axial angle conservation~\cite{AngularMultiplexing1981}, so the two can in principle be used at the same time, increasing the number of available parallel channels information can be transmitted through.

We have tested the radial memory effect with fibres as short as 20~cm and as long as 20~m without seeing any degradation, but it is likely that in fibres so long that nonlinear effects start to be significant the ring of extra intensity at the output would broaden gradually through mode coupling until all information on the input radial position is completely washed away.

The existence of the radial memory effect means that we can infer information on the radial distribution of the input light from the radial distribution of the (average) output intensity. If the radial distribution of the input light is simple enough, e.g., a discrete number of spots at known radii, it is possible to reconstruct it even from single-shot measurements. Exploiting this we showed that it is possible to reconstruct with a good fidelity which of 4 possible inputs was on (Figure~\ref{fig:Fig4}), which in principle could be used for spatial multiplexing.

Our simulations  indicate that the width of the bright ring of the radial memory effect in an ideal step-index fibre is independent of $r_0$, and is diffraction limited by the NA of the fibre, giving a ring of width $\sim\lambda/\text{NA}$ atop a relatively uniform background (see Fig.~\ref{fig:Fig2}(d)). This suggests that the number of annular channels exploiting the radial memory effect scales linearly with a step-index fibre's NA and core radius $a$, with an upper limit of $\sim a\text{NA}/\lambda$. Step-index fibres support a number of spatial fibre modes (at a single input polarisation) given by $\sim (\pi a \text{NA}/\lambda)^2$, and so the ratio of the number of annular channels to fibre modes (i.e., the full spatial channel capacity of the fibre) is given by $\sim\lambda/(\pi^2a\text{NA})$. This ratio is thus higher for fibres of smaller core radius, at least in the ideal case.

Experimentally, we observe that the width of the bright ring doesn't change significantly between fibres of core radius $a=25\,\mu$m and $a=100\,\mu$m, both of NA\,=\,0.22 (see Supplementary Information) in agreement with the above relations. However, in practice, the width of the bright ring is spread by mode coupling in real fibres, and our empirical evidence suggests that a focussed input at a larger radius produces sharper and easier to measure rings. The degree to which mode coupling acts to wash away the radial memory effect in real fibres will be dependent upon the length of the fibre and how contorted it is, but our experiments suggest it is a relatively robust phenomenon.

Interestingly, rings of excess output intensity have been observed before due to localised points of broadband fluorescent emission propagating through very short segments of fibre ($\sim$8\,mm in length)~\cite{rimoli2024demixing}. Broadband light innately averages away the fine details of the speckle, leaving a very clear signature of the radial memory effect. In ref.~\cite{rimoli2024demixing}, the existence of these correlations aided the demixing of fluorescence time traces used to record neuronal activity through multimode fibers. The bright rings were interpreted as an artifact of using a very short segment of fibre. In this work we have explained the underlying cause of this phenomenon, and have shown that the radial memory effect can persist through much longer lengths of fibre, opening up the range of possible applications.

\section{Methods}
\label{section:method}
\begin{figure}
    \centering
    \includegraphics[width=1\textwidth]{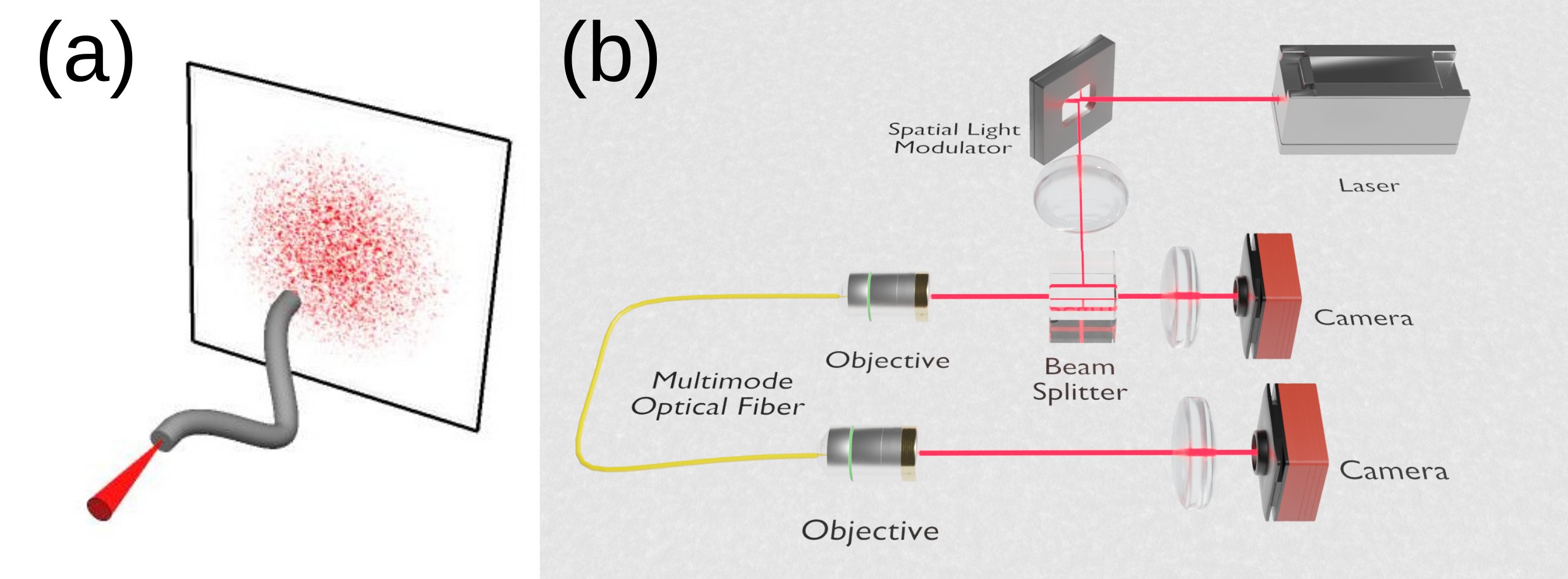}
    \caption{Experiment schematic. (a) In a multimode fibre, the incident wavefront will couple to the propagating modes, which each have a slightly different phase velocity. After propagating through the length of the fibre, the initial phase relation between these modes will be lost, and the superposition of the modes will now look like a speckle pattern. (b) Simplified schematic of the experimental apparatus. A digital micro-mirror device shapes incident laser light to generate one or more focussed spots at programmable locations on the input of a step-index multimode fibre. The cameras image the input and output facets of the fibre, allowing the intensity patterns at either end of the fibre to be compared.}
    \label{fig:Fig1}
\end{figure}

\begin{figure}
    \centering
    \includegraphics[width=1\textwidth]{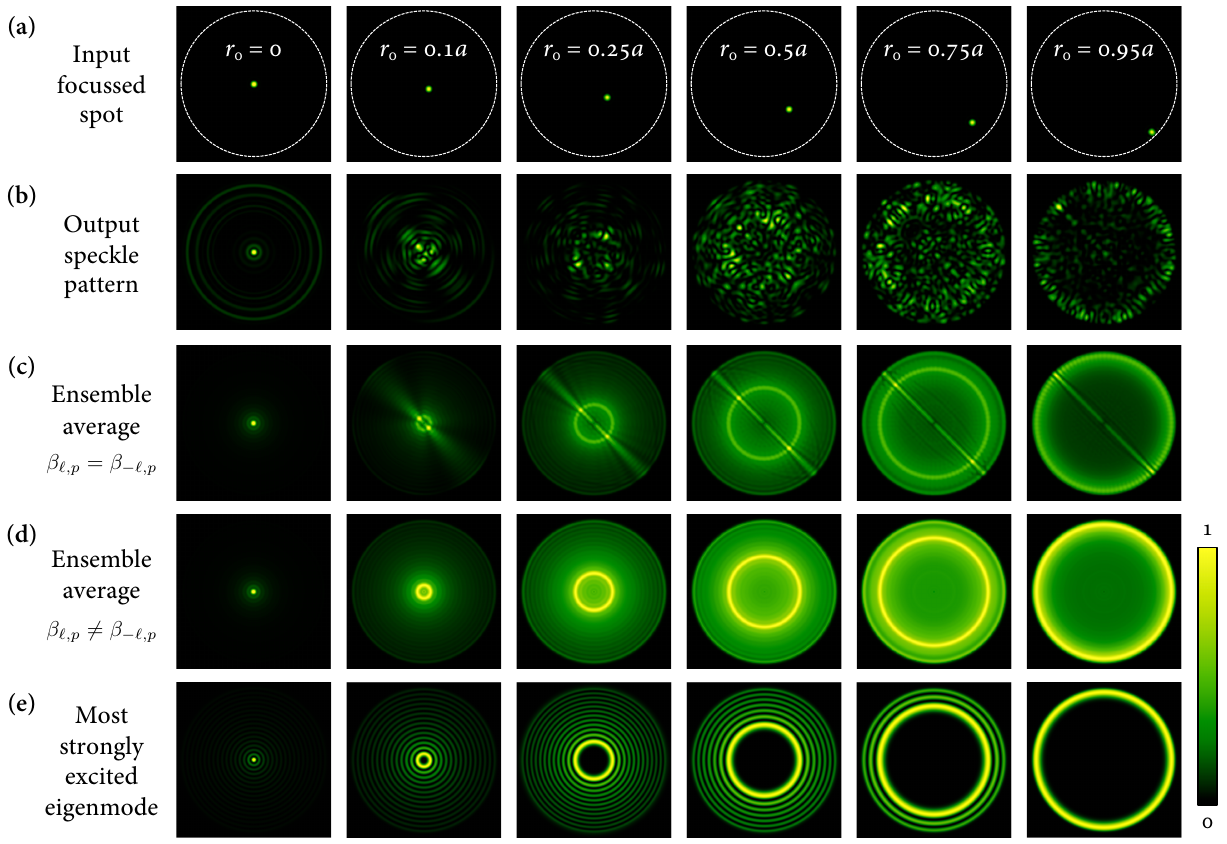}
    \caption{Numerical simulations of Eqn.~\ref{eq:intensity} and Eqn.~\ref{eq:avgint}. Here we simulate a step-index fibre of core radius $a=25\,\mu$m and NA = 0.22, illuminated by a wavelength of $\lambda=633\,$nm.
    (a) The six input focussed spot positions used for the numerical simulations. (b) Typical speckle patterns observed at the output end of an ideal straight fibre, when illuminated with the different input spot positions shown in (a). In all cases, a ring of brighter speckle is visible at the same radius as the input illumination. Here we simulate a fibre length of $Z$ = 2.4\,m. (c) Plot of Eqn.~\ref{eq:avgint} including all terms. The ring of excess intensity of radius $r_0$ (arising from the first term in Eqn.~\ref{eq:avgint}), and the diagonal line of excess intensity (arising from the second term in Eqn.~\ref{eq:avgint}) are both present. (d) Plot of Eqn.~\ref{eq:avgint} including only the first term on the right hand side (i.e., assuming that $\beta_{\ell,p}\neq\beta_{-\ell,p}$, as we expect to be the case in experiments). Here only the ring of excess intensity remains, while the diagonal line has disappeared. (e) The most strongly excited fibre mode for each of the input spot position, showing a high correlation with the position of the ring of excess intensity. All plots are individually normalised to the brightest point within each panel.}
    \label{fig:Fig2}
\end{figure}

\begin{figure}
    \centering
    \includegraphics[width=1\textwidth]{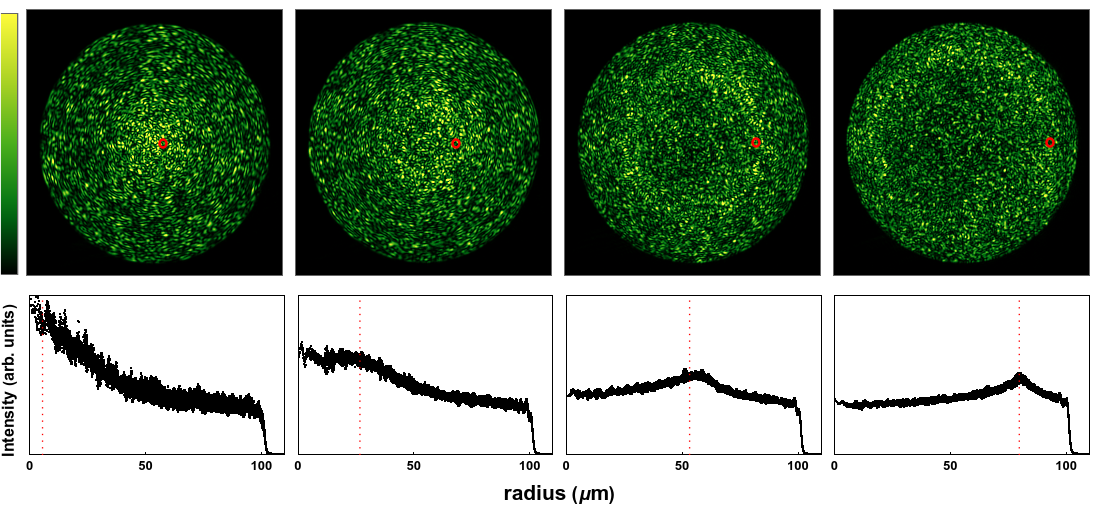}
    \caption{Experimental measurement of the radial memory effect. Top row: typical experimental speckle patterns at the end of a 20\,cm long, 200\,$\mu$m core diameter fibre for four different radial position of the input focus (red circle in the plots). A ring of excess intensity can be seen centered at the same radius of the input. Bottom row: projection on the radius of the speckle patterns in the top row, with the red dotted line showing the nominal radius of the input.}
    \label{fig:Fig3}
\end{figure}

\begin{figure}
    \centering
    \includegraphics[width=1\textwidth]{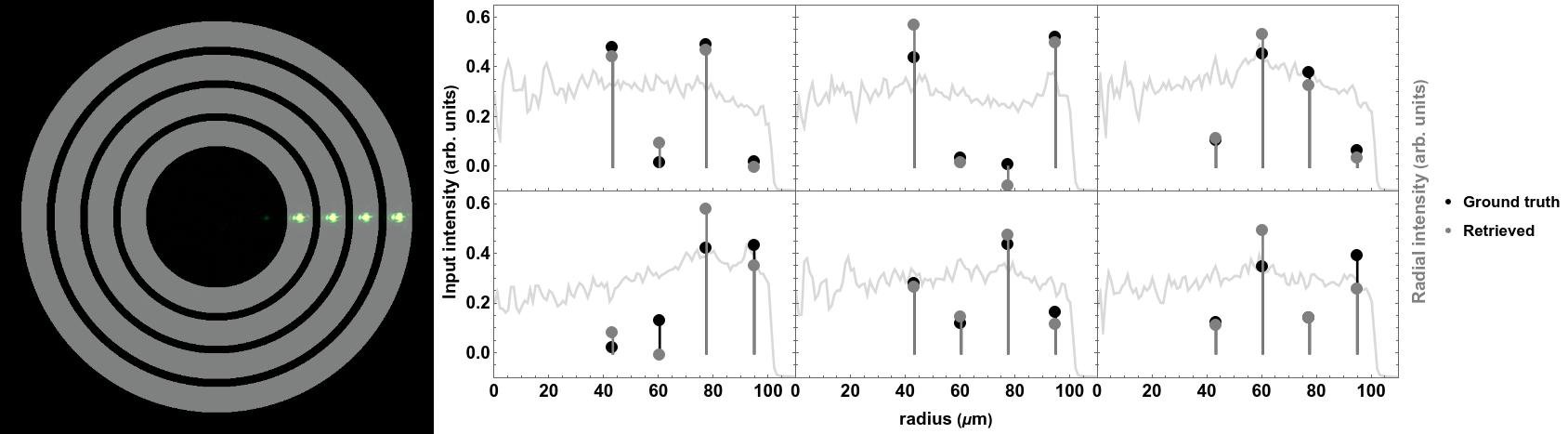}
    \caption{Spatial multiplexing using the radial memory effect. Left panel: The 4 spots used as inputs, and the 4 rings-shaped areas used to characterize the fibre response. Right panels: A typical selection of reconstructions (gray dots) compared with the ground truth (black dots) and the single-shot radial intensity distribution they were retrieved from (gray line).}
    \label{fig:Fig4}
\end{figure}

The optical system, depicted in Figure~\ref{fig:Fig1}, is designed to be able to illuminate an optical fibre with an arbitrary pattern and measure at the same time the light intensity on both facets of the fibre. We used a stabilized HeNe laser (Thorlabs HRS015B) with a wavelength of 632.99~nm. The beam is cleaned using a pinhole and then sent to a Digital Micromirror Device (DMD, a Vialux VX4100 with Texas Instruments DLP9500 VIS 1080p 0.95’’ chipset). The optical system is in a 4-f configuration where an adjustable aperture is used in the Fourier plane of the DMD to select out the +1-diffraction order of the DMD. After spatial filtering, the pattern generated in Fourier plane of the DMD is collimated and relayed to the back focal plane of a 10X Olympus objective. We note we also used a piezo actuated mirror in place of the DMD in early experiments. The objective images this pattern on to the multimode fibre’s front facet. Both the front and the distal facets of the fibre are imaged with identical optical systems with an overall 16x magnification to fill the camera’s active region with the imaged speckle disk. More details can be found in the Supplementary Information.

\section{Data availability statement}
The data that support the findings of this study are openly available at the following URL: doi.org/10.5281/zenodo.16760995

\section{Funding}
UK and JB acknowledge funding from QuantIC (EP/M01326X/1).  
RJK was supported by a Doctoral Training Partnership grant from the UK Engineering and Physical Sciences Research Council (EPSRC) (EP/T518049/1). DBP acknowledges financial support from the European Research Council (101170907). DBP and SARH acknowledge funding from the EPSRC (EP/Z535928/1).

\appendix

\section{Elastic scattering, speckle correlations and memory effects}
Every electric field with a finite energy are square-integrable, and thus can be represented as a vector in the $L^2$ Hilbert space (the same argument holds for magnetic fields, but since we are in the optical regime, we will neglect the magnetic response of the scattering media). If we limit ourselves to considering elastic scattering, then the relation between the incident and the scattered field must be linear, and we can write $E_{\text{out}}= \mathcal{S}\, E_{\text{in}}$, where $\mathcal{S}$ is a unitary operator (usually called the ``scattering matrix").\\
Looking at the scattering problem like this we can see that the effect of the (elastic) scattering is to rotate the electric field in the (high-dimensional) Hilbert space where it lives. Since rotations are always invertible, which is exactly what one tries to do when measuring $\mathcal{S}$, it means that no information is lost by scattering. This might look slightly unintuitive, as clearly a seemingly random speckle pattern appears to contain much less information than the unscattered field. The crucial point here is that speckle is indeed not random at all, merely complicated. If one had the key (i.e. a full characterization of $\mathcal{S}$) it would be possible to measure the scattered field and know exactly what the incident field was. On the other hand, after the scattering the information is easy to retrieve only in the abstract high-dimensional Hilbert space, but it is not directly accessible to us. In the space we inhabit, the information has been moved from the field profile to the speckle correlations.

If speckle was truly random (i.e., it contained no information) it would have no correlations, but since it must contain the same amount of information as the unscattered field, it must have correlations, and those correlations must actually contain all the information we are apparently missing. There is a large number of possible correlations one can investigate, but the simplest class are 2-point correlations, i.e., ``if I know the field at point $\mathbf{r}_1$, how much do I know about the field at point $\mathbf{r}_2$?".

To make things more quantitative we can define a correlation coefficient as
\begin{equation}
    C_{a,b,a^{\prime},b^{\prime}} = \frac{\left\langle \delta \mathcal{S}_{a,b} \, \delta \mathcal{S}_{a^{\prime},b^{\prime} }\right\rangle }{\left\langle \mathcal{S}_{a,b} \right\rangle\, \left\langle \mathcal{S}_{a^{\prime},b^{\prime}} \right\rangle} \, ,
\end{equation}
where $\left\langle . \right\rangle$ represents an ensemble average over the disorder realization (i.e., over all the possible scatterers' position), $\mathcal{S}_{a,b}$ is the matrix element of the scattering matrix $\mathcal{S}$ for input $a$ and output $b$ (this notation is completely agnostic with respect to which basis one chooses to describe the fields), and $\delta \mathcal{S}_{a,b} = \mathcal{S}_{a,b} - \left\langle \mathcal{S}_{a,b} \right\rangle$.

2-points speckle correlations have been extensively studied~\cite{Feng1988}, but their application to imaging and information retrieval is still an active field of research~\cite{Bertolotti2022}. The sub-class of 2-point speckle correlations that found the most applications is when $C$ doesn't depends independently on the two inputs and the two outputs, but only on $(a-a^{\prime})-(b-b^{\prime})$. These correlations are known as ``memory effects" and allow to predict how the output will change by knowing how the input was changed. The most famous example in optical fibres is the axial angle conservation in step-index multimode fibres, meaning that illuminating a fibre with a beam at an angle $\theta$ will result in an output largely concentrated in a cone of the same angle, which can be used for angular multiplexing~\cite{AngularMultiplexing1981}. Another, more recently discovered memory effect in fibres is the rotational memory effect, which connects the rotation of the input wavefront to the rotation of the output wavefront~\cite{amitonova2015rotational}, and has found application for imaging through a fibre~\cite{li2021memory}.

\section{Intensity profile theory}
To find the intensity profile at the distal end of the fibre we decompose the field into the propagating modes using the weakly guiding approximation
\begin{equation}\label{eqn:eigenmode}
\psi_{\ell,p}\left(r,\theta, z \right)= \frac{1}{N_{\ell,p}}\frac{J_{\ell}\left({u}_{\ell ,p}r/a\right)}{J_{\ell}\left(u_{\ell ,p}\right)} {e}^{{\rm{i}}\ell\theta}\, {e}^{{\rm{i}} \beta_{\ell,p} z},
\end{equation}
where $a$ is the radius of the fiber core, $\ell$ and $p$ enumerate the modes, $J_{\ell}$ is a Bessel function of the first kind of order $\ell$, $u_{\ell ,p}$ is the normalized transverse wave number, $\beta_{\ell,p}$ is the $z$ component of the wavevector, and $N_{\ell,p}$ is a normalization factor. If we excite the fibre at $z=0$ with a very small focus at $r=r_0$ and $\theta=\theta_0$ the coupling coefficients are given by
\begin{equation}
    C_{\ell,p} =\int^\pi_{-\pi}\int_0^a\frac{1}{r_0}\delta(r-r_0)\delta(\theta-\theta_0)\, {\psi }^*_{\ell ,p}\left(r,\theta\right)r\,dr\,d\theta = {\psi }^*_{\ell ,p}\left(r_0,\theta_0\right).
\end{equation}
As we are not interested in the fine detail of any particular output speckle pattern, we start by calculating the output intensity ${\langle I\left(r,\theta,r_0,\theta_0\right)\rangle_Z}$, averaged over a range of fibre lengths $Z$, when the fibre is illuminated by a focused point at radius $r_0$ and azimuthal angle $\theta_0$ on the input:
\begin{equation}\label{eq:IntAv}
\begin{aligned}
\langle I&\left(r,\theta,r_0,\theta_0\right)\rangle_Z = \left\langle \frac{1}{M}\left|\sum_{\ell,p}C_{\ell,p}\,{\psi }_{\ell ,p}\left(r,\theta, z\right)\right|^2 \right\rangle_Z =\\
&= \frac{1}{M\Delta}\int_{-\frac{\Delta}{2}}^{\frac{\Delta}{2}} {\rm d}z\left|\sum_{\ell=-L}^{L}\sum_{p=0}^{P(\ell)}\psi_{\ell,p}^{\ast}(r_0,\theta_0,0)\psi_{\ell,p}(r,\theta, z+Z) \right|^{2}
,
\end{aligned}
\end{equation}
where there are $M$ modes in total that can propagate within the fibre at this wavelength, and the averaging is taken over a change in length from $Z-\Delta/2$ to $Z+\Delta/2$.  The $z$--dependence within the integrand occurs only within the exponential, with the result of the integral equal to
\begin{equation}
    \frac{1}{\Delta}\int_{-\frac{\Delta}{2}}^{\frac{\Delta}{2}}{\rm d}z\,{\rm e}^{{\rm i}(\beta_{\ell,p}-\beta_{\ell',p'})z}=\frac{\sin\left((\beta_{\ell,p}-\beta_{\ell',p'})\frac{\Delta}{2}\right)}{(\beta_{\ell,p}-\beta_{\ell',p'})\frac{\Delta}{2}}\to\begin{cases}0&\beta_{\ell,p}\neq\beta_{\ell',p'}\\1&\beta_{\ell,p}=\beta_{\ell',p'}\end{cases}\label{eq:averaging}
\end{equation}
where in the final step of Eq. (\ref{eq:averaging}) we took the limit of a large range of fibre lengths, much larger than $\beta_{l,p}^{-1}$.  Applying the result (\ref{eq:averaging}) to the sum (\ref{eq:IntAv}),
\begin{equation}
\label{eq:final-average}
\begin{aligned}
    \langle I(r,\theta,Z)\rangle&=\frac{1}{M}\sum_{\ell=-L}^{L}\sum_{p=0}^{P(\ell)}|\psi_{\ell,p}(r_0,\theta_0)|^{2}|\psi_{\ell,p}(r,\theta)|^{2}\nonumber +\\
    &+\frac{2}{M}{\rm Re}\left[\sum_{\ell=1}^{L}\sum_{p=0}^{P(\ell)}\psi_{-\ell,p}(r_0,\theta_0)\,\psi_{\ell,p}^{\ast}(r_0,\theta_0)\,\psi_{-\ell,p}^{\ast}(r,\theta)\,\psi_{\ell,p}(r,\theta)\right] =\\
    &= \frac{1}{M}\sum_{\ell=-L}^{L}\sum_{p=0}^{P(\ell)}\frac{J^2_{\ell}(u_{\ell,p}r_0/a)J^2_{\ell}(u_{\ell,p}r/a)}{N_{\ell,p}^4J^{4}_{\ell}(u_{\ell,p})} +\\
    &+\frac{2}{M}\sum_{\ell=1}^{L}\sum_{p=0}^{P(\ell)}\frac{J^2_{\ell}(u_{\ell,p}r_0/a)J^2_{\ell}(u_{\ell,p}r/a)}{N_{\ell,p}^{4}J^{4}_{\ell}(u_{\ell,p})}\cos(2\ell(\theta-\theta_0))
\end{aligned}
\end{equation}
To obtain the above result we assumed that the only degeneracies in the propagation constants $\beta_{\ell,p}=\beta_{\ell',p'}$ occur when $p=p'$ and $\ell=\pm \ell'$, the term in the fourth line arising from the interference of positive and negative angular momenta. Note that the summation range in the second and fourth lines exclude $\ell=0$.  This prevents double counting of the zero angular momentum mode, which has a propagation constant $\beta_{0,p}$ that is not degenerate with any other mode. In the second line we also make use of the fact that the sum of a complex number $A$ and its complex conjugate $A^*$ is equal to twice the real part of $A$.

\section{Further measurements}

\subsection{The radial memory effect in a long fibre}

\begin{figure}[h]
    \centering
    \includegraphics[width=1\textwidth]{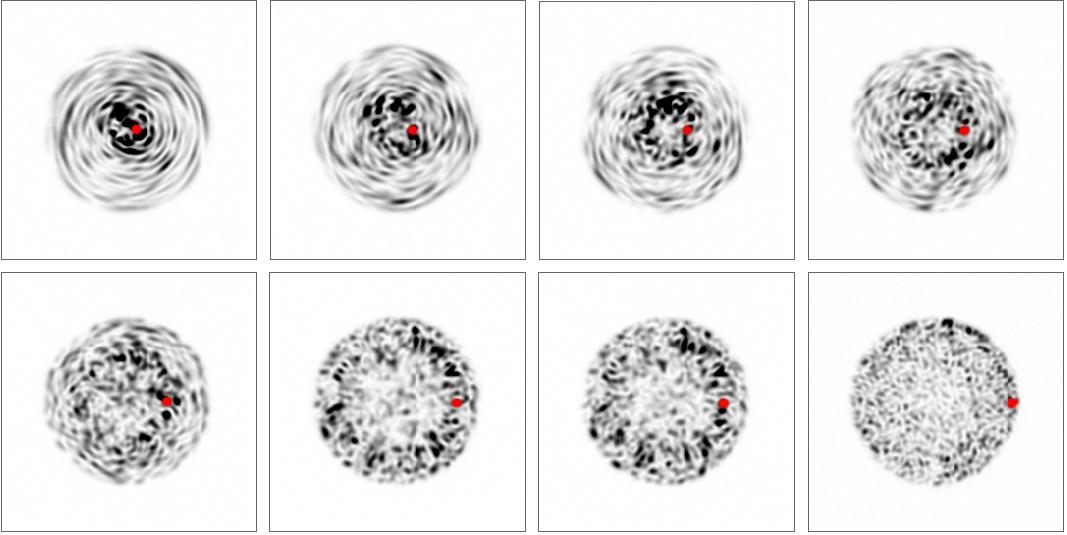}
    \caption{Typical experimental speckle patterns at the end of a 20~m long, 50$\mu$m core fibre for eight different radial position of the input focus (red dot in the plots). A ring of excess intensity can be seen centered at the same radius of the input.}
    \label{fig:long}
\end{figure}

In order to confirm that our results are not limited to short fibres, we repeated the measurements described in the main text on a 20\,m long, step-index fibre of NA=0.22 and core diameter 50\,$\mu$m. The results are shown in Figure~\ref{fig:long} and show that not only the ring of excess energy is clearly visible also for long fibres, but having a smaller core radius seems to make it even more visible.

\subsection{Input average}
\begin{figure}[th]
    \centering
    \includegraphics[width=1\textwidth]{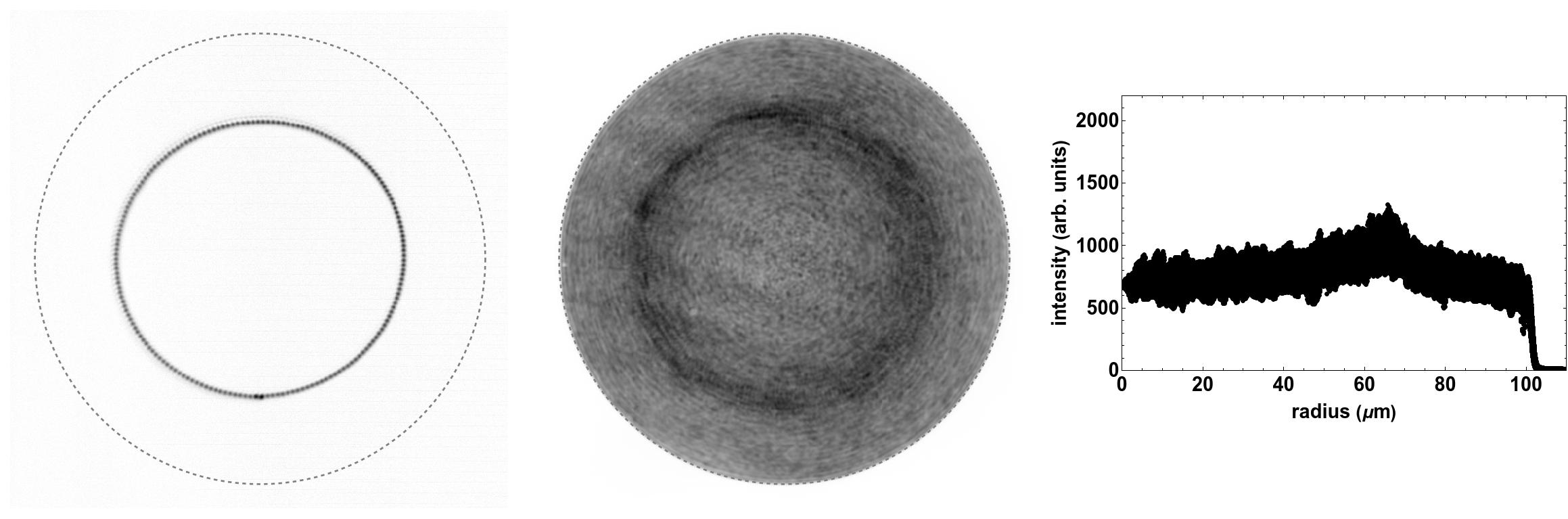}
    \caption{Left: 180 input spots at constant $r_0$ and uniformly spaced angles. Centre: Average over the input position, showing a clear ring of excess energy sitting on top of a flat background. The weak sinusoidal modulation is due to interference in the protective window of the camera, that acts like a weak Fabry-Perot. Right: projection on the radius of the average in the central panel, showing a well defined peak.}
    \label{fig:inputavg}
\end{figure}

As the position of the ring of excess energy only depends on $r_0$, one would reasonably expect that averaging over many input spot positions at the same $r_0$ would smooth out the fluctuations (i.e., the speckle pattern) leaving a clear ring. To test that we made 180 sequential illumination at fixed $r_0$ and uniformly spaced angles and averaged the resulting speckle patterns, as shown in Figure~\ref{fig:inputavg}.

\begin{figure}[tbh]
    \centering
    \includegraphics[width=1\textwidth]{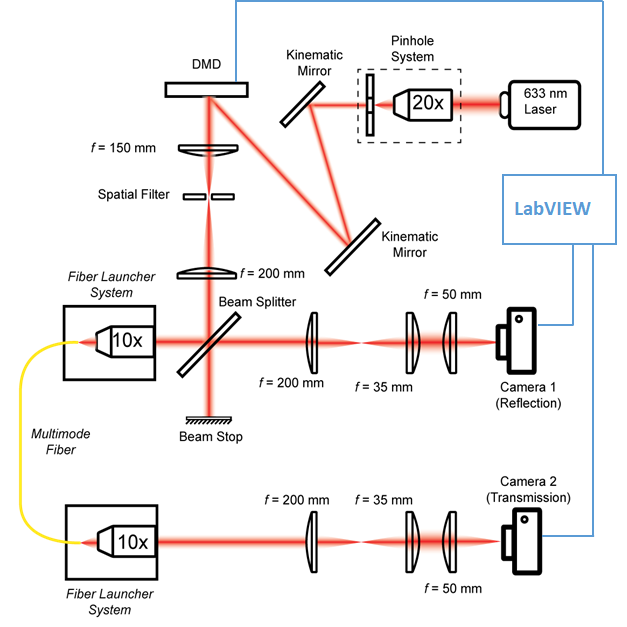}
    \caption{Detailed diagram of the experimental apparatus.}
    \label{fig:S2}
\end{figure}

\section{Experimental setup}
Figure~\ref{fig:S2} shows a detailed diagram of the experimental setup.
The system uses a DMD to arbitrarily shape the light on the input facet of a step-index multimode fibre, and simultaneously image both the input and output facet of the fibre. Input laser light is at a wavelength of 632.99\,nm from a HeNe laser (Thorlabs HRS015B). Light is produced in intensity stabilization mode of the laser and a small portion of the output is directed to the USB powermeter for monitoring the changes in the optical power output from the laser. After turning on the Thorlabs HRS015B laser, we wait for approximately 30 minutes for the laser to warm up and stabilize. The laser output passes through a simple pinhole system to produce a clean illumination spot. This spot is then relayed to the surface of the DMD.
The DMD used in this system is Vialux VX4100 with Texas Instruments DLP9500 VIS 1080p 0.95’’ chipset. The active region of the DMD is in horizontal configuration so that the light accepting corner of the DMD is at the bottom right corner. In this configuration, the input spot should arrive at the DMD with $22^o$ from the surface normal in the horizontal direction and $45^o$ from the surface normal in the perpendicular direction. When illuminated with such a spot, the main diffraction orders of the DMD will be produced in the surface normal direction of the DMD. This direction is parallel to the surface and along the mounting holes of the optical table.
The optical system is in 4-f configuration where an adjustable aperture is used in the Fourier plane to spatially filter out the +1-diffraction order of the DMD. After the spatial filtering, the pattern at the plane of the spatial filter is collimated and relayed to the back focal plane of a 10X Olympus objective lens. The objective images the pattern on to the multimode fibre’s front facet.
At the input side of the fibre, a beam splitter allows the light reflected from the input facet to be imaged onto a camera. This allows us to monitor the incident light pattern. The output facet of the fibre is also imaged onto a second camera. The light is collected by a second 10X Olympus objective lens. In both reflection and the transmission imaging arms, optical paths are infinity corrected and a magnification of about 16X is obtained to fill the camera’s active region with the imaged speckle disk.

\newpage

\bibliographystyle{unsrt}
\bibliography{bibliography}

\end{document}